# Studies of the Anomalous Hall Effect and Magnetic Structure of $Nd_2Mo_2O_7$
## - Test of the Chirality Mechanism-

Yukio Yasui, Taketomo Kageyama, Taketo Moyoshi, Minoru Soda, Masatoshi Sato and Kazuhisa Kakurai[1]

*Department of Physics, Nagoya University, Furo-cho, Chikusa-ku, Nagoya 464-8602*
[1]*Quantum Beam Science Directorate, Japan Atomic Energy agency, Tokai-mura, Naka-gun, Ibaraki, 319-1195*

**Abstract**

Neutron scattering studies have been carried out under the magnetic fields $H//[0\bar{1}1]$ and $H//[001]$ on a single crystal of $Nd_2Mo_2O_7$, whose Hall resistivity($\rho_H$) exhibits quite unusual $H$- and temperature($T$)-dependences. Material parameters such as the single ion anisotropies of the Mo- and Nd- moments and exchange coupling constants among the Mo-Mo, Mo-Nd and Nd-Nd moments, have been determined to reproduce various kinds of experimental data taken as a function of $H$ and $T$. For example, the neutron Bragg intensities, magnetization curves and the magnetic specific heats have been reproduced by the common parameters. By using the magnetic structure reproduced by these parameters, the spin chirality ($\chi$) of Mo spins or the fictitious magnetic flux $\Phi_{Mo}$ ($\propto \chi$) has been calculated as a function of $H$ and $T$ by using equation $\chi=<S_1 \cdot S_2 \times S_3>$, where the bracket means the statistical average. (Note that we do not use the equation $\chi=<S_1>\cdot<S_2>\times<S_3>$, because the local nature of the chirality should be correctly considered.) Comparing the calculated results with the observed $\rho_H$, we can conclude that the unusual behavior of $\rho_H$ cannot be understood consistently only by the chirality mechanism.

Keywords: pyrochlore molybdate, neutron scattering, magnetic structure, spin chirality, anomalous Hall resistivity

## §1. Introduction

$R_2Mo_2O_7$ (R=Y and rare earth elements) belongs to the pyrochlore series of compounds described by the general formula of $A_2B_2O_7$. The compounds have the face-centered cubic structure with the space group $Fd\bar{3}m$, in which two individual 3-dimensional networks are formed by corner-sharing $A_4$ and $B_4$ tetrahedra. For the magnetic moment at each corner of $A_4$ or $B_4$ tetrahedra the easy axis is often found along the line connecting the site with the center of gravity of the tetrahedron, suggesting that the moments align along four different but crystallographically equivalent directions, i.e., [111] and three other ones in the magnetically ordered phase. Then, non-trivial i.e., non-collinear or non-coplanar magnetic structure is often expected. In particular, for rare earth elements, the anisotropy-energy is generally large than for transition metal elements.

$R_2Mo_2O_7$ (R=Y and rare earth elements) exhibits an insulator to metal transition with increasing ionic radius of $R^{3+}$. $Nd_2Mo_2O_7$ is metallic and becomes ferromagnetic at the Curie temperature $T_c$=93 K.[1-3] With decreasing temperature ($T$), the magnetization ($M$) of $Nd_2Mo_2O_7$ begins to decrease below ~30 K. Neutron diffraction studies carried out by the authors' group[4-6] on a single crystal of $Nd_2Mo_2O_7$ revealed following characteristics of the magnetic structure in the $T$-region of 1.6 K $\leq T \leq T_C$. The ferromagnetic ordering at $T_C$=93 K is primarily associated with the Mo-moments, while the ordering of the Nd moments becomes significant below ~30 K. The net magnetization of the Nd-moments is antiparallel to that of the Mo-moments. The low $T$ magnetic structure is non-collinear for both the Mo- and Nd-moments: The Mo-moments align along the direction nearly parallel to the [001] or other equivalent axes, but slightly tilted from the direction by the internal magnetic field from the Nd-moments, while the Nd moments are along the directions almost parallel to their principal axes with their net magnetization being antiparallel to that of the Mo-moments. Because two of four Nd moments at the corners of a tetrahedron are inwards and the other two are outwards of the tetrahedron, the magnetic structure is called "two-in two-out" configuration. Detailed magnetic structure of $Nd_2Mo_2O_7$ is presented later.

Unusual behavior of the Hall resistivity ($\rho_H$) measured for single crystals of $Nd_2Mo_2O_7$ under the magnetic field ($H$) along [001], [110] or [111] direction has been reported by the authors' group,[7,8] and some of the data of $\rho_H$ measured for $H//$[111] are re-plotted against $H$ in Fig. 1, for example. The magnetization $M$ is also plotted in the inset. These data clearly show followings. In the $T$ region 30 K<$T$<$T_C$, where the magnetization is primarily from the Mo-moments, the $H$-dependence of $\rho_H$ is found to be described well by the well-known equation[9,10]

$$\rho_H = R_0 H + 4\pi R_s M, \quad (1)$$

by using the ordinary and anomalous Hall coefficients, $R_0$ and $R_s$ respectively, while below ~30 K, where the ordering of the Nd-moments is significant, eq. (1) no more works and another phenomenological equation

$$\rho_H = R_0 H + 4\pi R_s M_{Mo} + 4\pi R_s' M_{Nd}, \quad (2)$$

has to be used to describe the experimental results, where $M_{Mo}(M_{Nd})$ and $R_s(R_s')$ are the net magnetization and the anomalous Hall coefficient of the Mo (Nd)-moments, respectively. Solid lines in Fig.1 are the results of the fittings carried out by eq. (2). The sign of $R_s$ and $R_s'$ are found to be positive and negative, respectively ($M_{Mo}$ is defined to be positive and $M_{Nd}$ is negative if $M_{Nd}$ is antiparallel to $M_{Mo}$.). Effects of impurity-doping on $\rho_H$ have also been studied by the authors' group for single crystal samples of $Nd_2Mo_{2-x}Ti_xO_7$ ($x$=0.1 and $x$=0.3).[11] In the studies, we have found that with increasing Ti-doping the sign change of $R_s$ takes place with the sign of $R_s$' unchanged. These results indicate that two components are necessary to describe the $\rho_H$ behavior of $Nd_2Mo_2O_7$. Another interesting point is that $R_s$ and $R_s$' do not exhibit significant decrease as $T$ approaches zero, which seems to be rather different from the behavior of ordinary ferromagnets.[9,10]

In eq. (1), which can be applied for ordinary ferromagnets, the first term $R_0B$ is due to the effect of the Lorentz force acting on the conduction electrons moving in the magnetic field, where the effective magnetic field $B$ is equal to $H+4\pi(1-N)M$, with $N$ being the demagnetization coefficient ($R_0B \sim R_0H$ at $N \sim 1$). As for the origin(s) of the second term or the anomalous Hall effect, there are various kinds of theoretical studies, which consider both intrinsic and extrinsic mechanisms. Karplus and Luttinger[12] proposed that the existence of a term in the velocity operator proportional to the gradient of any electrical potential acting on the conduction electrons and to a contribution acting like a magnetic field. The latter has been identified with a Berry phase mechanism,[13-15] which is derived from the topological properties of Bloch states in a periodic potential of the crystal, moving in $k$-space under the influence of the combined effect of spin-orbit interaction and a ferromagnetic polarization. There exist even in the absence of any impurity scattering and therefore it is intrinsic mechanism. Impurities can also contribute to the anomalous Hall effect, where two mechanisms exist. One was reported by Smit[16] as the skew-scattering mechanism, where the Hall coefficient $R_s$ is expected to be proportional to the electrical resistivity $\rho$. Another was reported by Berger[17] as side-jump mechanism, where $R_s$ is proportional to $\rho^2$.

Regardless of what mechanism is responsible, $\rho_H$ depend on the spins of the conduction electrons through the spin-orbit coupling. For $Nd_2Mo_2O_7$, the anomalous component of $\rho_H$ has to be described by the Mo-spins. However, the anomalous component of the Hall resistivity of the Nd-moments ($4\pi R_s'M_{Nd}$) cannot be described by such mechanisms, because the $Nd^{3+}$ ions of $Nd_2Mo_2O_7$ do not have conduction electrons. The absolute values of $4\pi R_s'M_{Nd}$ is much larger than that of $4\pi R_0M_{Nd}$ (the values of $|4\pi R_s'M_{Nd}/4\pi R_0M_{Nd}|$ are equal to be ~ 45 for $H$//[001] and ~ 200 for $H$//[111], respectively) at $T$=5 K, indicating that the $4\pi R_s'M_{Nd}$ cannot be explained by the ordinary Hall effect, that is, the internal magnetic field generated by the Nd-moments cannot generate the Lorentz force on the $Mo^{4+}$ conduction electrons large enough to explain the magnitude of the anomalous term. It is interesting to investigate the mechanism or origin(s) of the unusual component of the term $4\pi R_s'M_{Nd}$ in eq. (2).

A proposal has been made that the ordering of the spin chirality ($\chi$) contributes to the Hall resistivity, where $\chi$ is locally defined as $\chi \equiv S_1 \cdot S_2 \times S_3$ for three spins $S_1$, $S_2$ and $S_3$.[18] This mechanism can be considered a special case of the Berry phase mechanism. It adapts the strong Hund's coupling between the localized spins and those of conduction-electrons. From this theory, the anomalous Hall conductivity $\sigma_H(=\rho_H/\rho^2)$ is proportional to the fictitious magnetic flux ($\Phi$) induced by the spin chirality $\chi$. Taguchi et al.[19] proposed that the spin chirality or the fictitious magnetic flux of the Mo-moments is playing an important role in determining the unusual behavior of the Hall resistivity of $Nd_2Mo_2O_7$. They also insist that the sign change of the Hall resistivity with increasing $H$ along the [111] direction is the evidence of the Hall resistivity induced by the chirality mechanism.[20] However, their discussion of the anomalous Hall resistivity by the chirality mechanism is made without a precise understanding of the magnetic structure of the Mo- and Nd-moments necessary to estimate the spin chirality $\chi$ ($\equiv S_1 \cdot S_2 \times S_3$). Based on the magnetic structure obtained by the neutron measurements, the authors' group has reported that the chirality mechanism does not consistently explain the observed behavior of $\rho_H$.[4-6] It is important for clarifying the relationship between the chirality mechanism and the unusual behavior of the Hall resistivity, to more precisely understand the magnetic structure of $Nd_2Mo_2O_7$, particularly to understand the $T$- and $H$-dependences of the magnetic structure.

In the present work, neutron scattering studies have been carried out on a single crystal of $Nd_2Mo_2O_7$ in the applied magnetic fields $H$//[0$\bar{1}$1] and $H$//[001] up to 5.7 T. The inelastic neutron scattering studies using by the cold neutrons have also been carried out at $T$=1.5 K. We have determined the material parameters such as the single ion anisotropies of the Mo- and Nd-moments and exchange coupling constants among the Mo-Mo, Mo-Nd and Nd-Nd magnetic moments to simultaneously reproduce the neutron data, magnetization curves and the behavior of the magnetic specific heat, where we have to consider not only the nearest neighbor Nd-Nd interaction but also the second nearest neighbor one. It indicates that the RKKY interaction has a certain role. The spin chirality and the fictitious magnetic flux contributed from the Mo-moments ($\Phi_{Mo}$) have been calculated as a function of $H$ and $T$ by using the obtained material parameters determined in the present study, where $\Phi_{Mo}$ is estimated from the statistically averaged value of the fictitious flux generated by the local chirality, that is, the relation $\Phi_{Mo} \propto \chi=<S_1 \cdot S_2 \times S_3>$ is used. If we use the one $\Phi_{Mo} \propto \chi=<S_1>\cdot<S_2> \times <S_3>$, we may not necessarily be able to obtain meaningful result, because the local nature of the chirality is not correctly considered. (The bracket means the statistical average.) By comparing the Hall resistivity $\rho_H$ deduced from the fictitious magnetic flux with the experimentally observed data, we have found that the unusual behavior of the Hall resistivity of $Nd_2Mo_2O_7$ cannot consistently be understood by the chirality mechanism.

§2. Experiments

Crystals of $Nd_2Mo_2O_7$ used in the previous and present studies were grown by the floating zone method under the Ar-atmosphere. The detailed description on the experimental data of the magnetization $M$ and the Hall resistivity $\rho_H$ can be found in the previous papers.[7,8,11] (Some of the data can also be found in Fig. 1.) Neutron measurements of the Bragg reflection intensity were carried out by using the triple axis spectrometer HQR(T1-1) and PONTA(5G) installed at JRR-3M of JAEA in Tokai, where the double axis condition was adopted. Almost all data taken at T1-1 can also be found in ref. 6. New data were taken at 5G for a crystal different from the one used at T1-1, where the horizontal collimations were 40'-40'-80' and the neutron wavelength was 2.350 Å. In the measurements, the crystal was oriented with the [100] and [010] axes in the scattering plane. (In the previous measurements at T1-1, the crystal was set with the [011] and [100] axes in the scattering plane.) The magnetic field was applied by using a superconducting magnet along the vertical direction. The 002 reflection of Pyrolytic graphite (PG) was used for both the monochromator and the analyzer. PG filters were placed after the second collimator and after the sample to suppress the higher-order contamination. In the analyses of the data, the isotropic magnetic form factors for $Nd^{3+}$ and $Mo^{4+}$ were used,[22,23] and the absorption and extinction corrections were made, where details of the extinction corrections have already been reported in ref. 4.

Inelastic neutron scattering was carried out at HER(C1-1) of the cold guide installed at JRR-3M of JAERI in Tokai, for the same sample as used at T1-1 by using the triple axis spectrometer. The 002 reflections were used for both the monochromator and the analyzer. The final neutron energy $E_f$ was fixed at 2.424 meV, where the energy resolution $\Delta E$ (full width at half maximum) determined by the incoherent nuclear scattering from a vanadium rod was 0.052 meV. To remove the higher-order neutrons, Be and PG filters were placed in front of the sample and a Be filter was placed after the sample. The crystal was oriented with the [1$\bar{1}$0] direction vertical, where the [110] and [001] axes were in the scattering plane.

§3. Experimental Results

Examples of the energy scan profiles of the magnetic scattering taken at C1-1 for the $Nd_2Mo_2O_7$ sample are shown at 1.5 K in Fig. 2, where well-defined magnetic excitations can be found at the transfer energy $E \sim 0.5$ meV. The characteristics of the $T$- and $Q$-dependences of these profiles indicate that these peaks are derived from magnetic scattering. The profiles are analyzed in the region 0.26 meV$\leq E(=\hbar\omega) \leq 2.0$ meV by using the following expression of the dynamical structure factor

$$S(\boldsymbol{Q},\omega) \propto \frac{1}{1-e^{-\beta\omega}} \cdot \chi(\boldsymbol{Q}) \cdot \frac{\Gamma(\boldsymbol{Q})\cdot\omega}{(\omega-\omega_Q)^2 + \Gamma(\boldsymbol{Q})^2}, \quad (3)$$

where $\omega_Q$ and $\Gamma(\boldsymbol{Q})$ are the energy and the energy broadening of the magnetic excitations, respectively. The static susceptibility is denoted as $\chi(\boldsymbol{Q})$ and $\beta=(k_BT)^{-1}$. The results of the fittings by eq. (3) to the experimental data are shown in Fig. 2 by the solid lines, where the fittings are satisfactory. Inset of Fig. 2 shows the example of the dispersion relation along $(h,h,2-h)$ obtained by the analyses, where the excitation is found to be dispersionless, indicating that the observed inelastic peaks can be considered to be from the local excitation of the Nd-moments with strong axial (Ising-like) anisotropy. It is consistent with the obtained magnetic structure at $T$=1.6K, where the Nd moments are along the directions almost parallel to their principal axes.[4-6] The values of $\omega_Q$ (=0.48 meV =5.6 K) correspond to that of $2\boldsymbol{\mu}_{Nd}\cdot\langle\boldsymbol{H}_{int}\rangle$, where $\boldsymbol{\mu}_{Nd}$ is the Nd-moments and $\langle\boldsymbol{H}_{int}\rangle$ is the averaged internal magnetic field at $\boldsymbol{\mu}_{Nd}$ sites coming from the surrounding Mo- and Nd-moments. The resolution correction was not made, because the resolution width ($\Delta E$ =0.052 meV) is much smaller than the value of $\Gamma(\boldsymbol{Q})$ (~ 0.29 meV), where the obtained $\Gamma(\boldsymbol{Q})$ is $\boldsymbol{Q}$-independent. The existence of the broadening $\Gamma(\boldsymbol{Q})$ much larger than the resolution width, indicates that the internal magnetic field has a distribution at Nd sites.

In the fields $\boldsymbol{H}$, along the directions parallel to [0$\bar{1}$1] and [001], the neutron scattering intensities have been measured at $\boldsymbol{Q}$=(h,k,k) and (h,k,0) in the reciprocal space, respectively, at $H$=0, $H$=3 T and $H$=5 T. Among these $\boldsymbol{Q}$-points, those with half integer values of $h$ and $k$ are included at 1.6 K (<$T_C$) and ~120K (>$T_C$). However, we have not observed Bragg reflections at $\boldsymbol{Q}$-points except the nuclear Bragg points. The scattering intensities at several nuclear Bragg points increase with decreasing $T$ through the Curie temperature. The intensities also depend on the values of the external magnetic field $H$ applied along the [0$\bar{1}$1] and [001] axes. Examples of the observed $\omega$-scan (sample-angle scan) profiles of several reflections taken under the magnetic filed along $\boldsymbol{H}$//[001] are shown in Figs. 3(a)-3(c), where the data at 120 K (>$T_C$) are the contribution from the nuclear reflection. Because the 200 and 600 nuclear reflections have no scattering intensity, the peaks observed at low temperatures below $T_C$ can be considered to be from the magnetic scattering. The profile widths of all reflections observed at 1.6 K are equal to those of the instrumental resolution within the experimental error bars.

The system have the face-centered cubic structure with the space group F$d\bar{3}m$, in which Nd and Mo atoms fully occupy the sites 16d(Nd) and 16c(Mo), and the crystal structure consists of two networks individually formed by corner sharing tetrahedra of $Mo_4$ or $Nd_4$. In this face-centered cubic cell, each one of four primitive cells has a structure unit consisting of a $Mo_4$-tetrahedron and a $Nd_4$-one. In the measurements of neutron scattering in the zero- and finite magnetic fields, the magnetic reflections have been observed only at nuclear Bragg points with respect to the f.c.c. unit cell. This result indicates that the magnetic primitive cell of $Nd_2Mo_2O_7$ has the same size as the chemical primitive cell. Then, to determine the magnetic structure of the present system, we are just required to find the moment arrangements within each one of $Mo_4$- and $Nd_4$-tetrahedra (or within a magnetic primitive cell).

The integrated intensities of the magnetic scattering ($I_{mag}$) of $Nd_2Mo_2O_7$ of the 200, 400, 600 and 240 reflections are shown at 1.6K against the magnetic field along $\boldsymbol{H}$//[001] in Figs. 4(a)-4(d), respectively, where the data were taken after zero field cooling (ZFC) by scanning $H$ stepwise up to 5.7 T ($H$-increase) and then to zero ($H$-decrease). Hysteretic behavior of the $I_{mag}$ –$H$ curves has been observed in the region of $H \leq$1.5 T. It is due to the hysteretic domain motion. $I_{mag}$ was obtained by taking the differences between the integrated intensities below and above $T_C$, where the absorption- and extinction-corrections were considered. Effects of the ambiguity of the extinction correction do not bring about essential changes of the results of the present study, because only the data at the Bragg points where the extinction effect is not serious (or scattering intensity is relatively weak) are mainly considered in the magnetic structure analysis.

In Figs. 5(a)-5(e), the $I_{mag}$-$H$ curves of the 111, 200, 022, 311 and 400 reflections taken at 1.6K under the magnetic field along $\boldsymbol{H}$//[0$\bar{1}$1] are reproduced from the previous report.[6] Note that the data were taken after zero field cooling (ZFC) by scanning $H$ stepwise up to 5.7 T ($H$-increase) and then to zero ($H$-decrease).

For the 200, 600 and 240 reflections under the field $H$//[001], the $I_{mag}$-$H$ curves have the minimum at $H$=3.0 T and the symmetric behavior of the $I_{mag}$-$H$ curves is observed with respect to the point $H$=3.0 T, except the data points obtained for the sample with the multi-magnetic-domain. The similar behavior is also observed the $I_{mag}$-$H$ curves of the 200 and 022 reflections for $H$//[0$\bar{1}$1]. The magnetic reflections with the asymmetric behavior of the $I_{mag}$-$H$ curves have interference term between the Mo-and Nd- moments. These results indicate that the average internal magnetic field at the Nd sites from the Mo-moments ($\langle H_{int}^{Nd \leftarrow Mo} \rangle$) is estimated to be 3.0 T and $\langle H_{int}^{Nd \leftarrow Mo} \rangle$ can be canceled out by applying the external magnetic field.

In Fig. 6, the magnetic scattering intensities of the 200 reflection ($I_{200}$) of $Nd_2Mo_2O_7$ is shown against $T$ at $H$=0. The appearance of the magnetic scattering intensity of the 200 reflection is due to the existence of the transverse component (the component perpendicular to the net magnetization of the Mo- and Nd-moments) of the ordered moments. From the magnetic structure analysis at $H$=0, the transverse component is mainly contributed from the Nd-moments. Then, the values of $I_{200}$ represent the ordered component of the Nd-moments. Preliminary analyses of the $I_{200}$-$T$ curve have been carried out by using the expression, $I_{200} \propto \{Tanh(2\Delta/T)\}^2$, which is generally applied to the simple two energy level system with energy gap $2\Delta$. From the fitting, $2\Delta$ is estimated to be 5.6 K, which is consistent to that of $\omega_Q$ (= $2\mu_{Nd} \cdot \langle H_{int} \rangle$). The characteristic energy estimated from the behavior of the $I_{200}$-$T$ curve is equal to 2.8 K (=$\Delta = \mu_{Nd} \cdot \langle H_{int} \rangle$), not $T^* \sim 40$ K and ~50 K, suggested in refs. 20 and 25, respectively.

In Figs. 7(a)-7(c), the Nd-components of the magnetizations ($M_{Nd}$) are shown at various temperatures against the magnetic field $H$ applied along [001], [110] and [111], respectively. The values of the $M_{Nd}$ are estimated by subtracting the magnetizations $M$ observed at 40 K from those at the lower temperatures, that is, the magnetization at 40 K is regarded as the Mo-component of the magnetization. Parts of the raw data of the magnetization curves were reported by authors' group in refs. 6-8. The sign change of the $M_{Nd}$ is observed at 3 T with increasing $H$ along all directions, which is consistent to the behavior of the magnetic scattering intensity. These results indicate that the direction of the net magnetization of the Nd-moments is reversed to the direction parallel to the Mo-moments at 3 T (=$\langle H_{int}^{Nd \leftarrow Mo} \rangle$) by the external magnetic field.

Insets of Figs. 8(a) and 8(b), the magnetic contribution of the specific heat ($C_{mag}$) divided by $T$ of $Nd_2Mo_2O_7$ is shown against $T$ under various magnetic fields $H$//[001] and $H$//[111], respectively. The experimentally obtained $C_{mag}/T$ data of $Nd_2Mo_2O_7$ is estimated by subtracting the $C/T$ data of $Nd_2Zr_2O_7$ (open circles) from those of $Nd_2Mo_2O_7$, where $C$ is specific heat. The $C_{mag}/T$ data of $Nd_2Mo_2O_7$ shown by the solid circles were reported previously by authors' group in refs. 23 and the data shown by the lines were reported in refs. 24. In pyrochlore compound $Nd_2Zr_2O_7$, the Curie-Weiss temperature estimated from the magnetic susceptibility measurements is smaller than 1 K[26], indicating that the $C/T$ data in the $T$-region 7 K$\leq T$ is considered to be contribution from phonons. In the analyses, the phonon contribution of $Nd_2Mo_2O_7$ is estimated by the behavior of the $C/T$ data of $Nd_2Zr_2O_7$ in the $T$-region 7 K$\leq T$ and by extrapolating the $C/T$ data of $Nd_2Zr_2O_7$ to the $T$-region $T$<7 K. The peaks of the $C_{mag}/T$–$T$ curves in the $T$-region $T \leq 15$ K are due to the magnetic contribution of the Nd-moments with Ising-like anisotropy. The behavior of the $C_{mag}/T$ –$T$ curves depends on the Nd-Mo and Nd-Nd interactions and the applied magnetic field. We tried to fit the $C_{mag}/T$ data by the Schottky type expression, which is generally applied to simple two energy level systems, but the broadened peak cannot be reproduced. The width of the peak of the $C_{mag}/T$ –$T$ curve indicates that the internal magnetic field at Nd sites has a distribution, which is consistent to the existence of the finite $\Gamma(Q)$ obtained by the inelastic neutron scattering.

§4. Analyses and Discussion

In the previous chapter, we have shown the results of the elastic and inelastic neutron scattering as well as the behaviors of the magnetization and magnetic contribution of the specific heat of $Nd_2Mo_2O_7$. Here, we analyze these data by using common material parameters. The effective Hamiltonian of the localized Mo - and Nd- moments has the following expression:

$$H = -\sum_{<m,n>} J_{Mo-Mo} S_m^{Mo} \cdot S_n^{Mo} - \sum_{<m,r>} J_{Mo-Nd} S_m^{Mo} \cdot S_r^{Nd}$$
$$- \sum_{<r,s>} J_{Nd-Nd} S_r^{Nd} \cdot S_s^{Nd} + \sum_m g_{Mo} \mu_B \langle f \rangle S_m^{Mo} \cdot H$$
$$+ \sum_s g_{Nd} \mu_B S_s^{Nd} \cdot H \quad (4)$$

, where $J_{Mo-Mo}$, $J_{Mo-Nd}$ and $J_{Nd-Nd}$ are the exchange coupling constants between Mo-Mo, Mo-Nd and Nd-Nd, respectively. $S_m^{Mo}$ and $S_s^{Nd}$ are spins at the Mo $m$-th site and at the Nd $s$-th one, respectively. $\langle f \rangle$ is a ratio of the number of the ordered Mo-moments to spin number nominally estimated from $Mo^{4+}$. For the $g$-factors of the Mo- and Nd-moments, we have used $g_{Mo}$ =2 and $g_{Nd}$ =0.727, respectively. The sum of the Mo-Mo interaction <$m,n$> and the Mo-Nd one <$m,r$> are over all pairs of neighboring spins. For the sum of the Nd-Nd interaction <$r,s$>, we have to consider not only pairs of neighboring (NN) sites but also all the pairs of the second neighbor (2nd) sites, where the former exchange coupling constants is $J_{Nd-Nd}^{NN}$ and the latter one is $J_{Nd-Nd}^{2nd}$. (Here, the second neighbor means that two Nd atoms cannot be connected with one edge of the $Nd_4$ tetrahedra but connected by two edges of the $Nd_4$ tetrahedra.) From the results of the neutron scattering studies and another measurements, the single ion anisotropy of the Mo-moments is found to be small, whereas the Nd-moment has strong axial anisotropy (Ising-like). We have chosen the material parameters as $J_{Mo-Mo}$, $J_{Mo-Nd}$, $J_{Nd-Nd}^{NN}$, $J_{Nd-Nd}^{2nd}$, $S^{Mo}$ and $S^{Nd}$ to reproduce the experimentally obtained data. From the magnetization data and the obtained magnetic structure at $H$=0,[4-8] the values of the saturation moments of Mo is estimated to be 1.30($\pm 0.10$) $\mu_B$ (= $g_{Mo} \mu_B \langle f \rangle S^{Mo}$), indicating $\langle f \rangle$ =0.65 and the absolute value of $S^{Mo}$ =1 ($Mo^{4+}$). The $T$-dependences of the values of the ordered Mo-moments and $\langle f \rangle$ are estimated from the $T$-dependence of the magnetic scattering intensity of the 111 reflection reported by the authors' group.[4)] $J_{\text{Mo–Mo}}$ is estimated to be 9.8 K (ferromagnetic) from the value of the Curie temperature $T_C$=93 K, which is primarily due to the ordering of the Mo-moments. In order to determine the other parameters, we have fitted the following expression to the observed data of $M_{\text{Nd}}(\boldsymbol{H},T)$, $C_{\text{mag}}(\boldsymbol{H},T)$ and the magnetic scattering intensity $I_{\text{mag}}(\boldsymbol{Q},\boldsymbol{H},T)$:

$$M_{\text{Nd}}(\boldsymbol{H},T) = \frac{\sum_i \left[ -g_{\text{Nd}}\mu_B (\sum_s \boldsymbol{S}^{\text{Nd}}_{s,i} \cdot \hat{\boldsymbol{H}}) \times \exp(-\varepsilon_i/(k_BT)) \right]}{\sum_i \exp(-\varepsilon_i/(k_BT))}, \quad (5)$$

$$C_{\text{mag}}(\boldsymbol{H},T) = \frac{\partial}{\partial T}\left[ \frac{\sum_i \varepsilon_i \cdot \exp(-\varepsilon_i/(k_BT))}{\sum_i \exp(-\varepsilon_i/(k_BT))} \right] \quad \text{and} \quad (6)$$

$$I_{\text{mag}}(\boldsymbol{Q},\boldsymbol{H},T) \propto \left| \left\{ \sum_i \left[ \sum_m p f_{\text{Mo}}(\boldsymbol{Q}) g_{\text{Mo}} \langle f \rangle \boldsymbol{S}^{\text{Mo}}_{m,i\perp} \times \exp(-i\boldsymbol{Q}\cdot\boldsymbol{R}^{\text{Mo}}_m) \right. \right. \right.$$
$$\left. + \sum_s p f_{\text{Nd}}(\boldsymbol{Q}) g_{\text{Nd}} \boldsymbol{S}^{\text{Nd}}_{s,i\perp} \times \exp(-i\boldsymbol{Q}\cdot\boldsymbol{R}^{\text{Nd}}_s) \right]$$
$$\left. \left. \times \exp[-\varepsilon_i/(k_BT)] \right\} \middle/ \left[ \sum_i \exp(-\varepsilon_i/(k_BT)) \right] \right|^2, \quad (7)$$

where $f_{\text{Mo}}(\boldsymbol{Q})$ and $f_{\text{Nd}}(\boldsymbol{Q})$ correspond to the magnetic form factor of the Mo- and Nd-moments, respectively, and $p$ is a magnetic scattering amplitude ($p$=0.269×10$^{-12}$ cm). $\hat{\boldsymbol{H}}$ indicates the unit vector of the external field direction. $\boldsymbol{R}^{\text{Mo}}_m$ and $\boldsymbol{R}^{\text{Nd}}_s$ are the position of the Mo $m$-th and the Nd $s$-th sites, respectively. The absolute values of $I_{\text{mag}}(\boldsymbol{Q},\boldsymbol{H},T)$ are estimated by comparing the magnetic scattering intensities with those of the nuclear scattering. Calculations of the above quantities have been carried out by considering the 16 Mo- and 16 Nd-moments within an f.c.c. unit cell under the periodic boundary condition. All magnetic configurations of the Nd moments in the unit cell have been considered, where the Nd-moment are pinned along the lines, which connect the Nd sites with the center of gravity of the Nd$_4$ tetrahedra, and therefore have the Ising-like anisotropy. The value of $\varepsilon_i$, the energy of the $i$-th ($i$ =1–65536(=2$^{16}$)) magnetic configuration can be calculated by using the above Hamiltonian. The probability of the $i$-th magnetic configuration can be calculated as a function of $\varepsilon_i$ by using the canonical distribution. $\boldsymbol{S}^{\text{Mo}}_{m,i\perp}$ ( $\boldsymbol{S}^{\text{Nd}}_{s,i\perp}$ ) represents the spin components of $\boldsymbol{S}^{\text{Mo}}_{m,i}$ ( $\boldsymbol{S}^{\text{Nd}}_{s,i}$ ) perpendicular to the scattering vector $\boldsymbol{Q}$, for the $i$-th magnetic configuration. As the results of the fittings, we have chosen the values: $S^{\text{Nd}}$ =2.65 ( $g_{\text{Nd}}\mu_B S^{\text{Nd}}$=1.93 $\mu_B$), $J_{\text{Mo–Nd}}$ = –0.25 K (antiferromagnetic), $J^{\text{NN}}_{\text{Nd–Nd}}$ =0.29K (ferromagnetic), $J^{\text{2nd}}_{\text{Nd–Nd}}$ = $-1/6 \times J^{\text{NN}}_{\text{Nd–Nd}}$ (antiferromagnetic). The consideration of $J^{\text{2nd}}_{\text{Nd–Nd}}$ is necessary to reproduce the $C_{\text{mag}}/T$–$T$ curves and different sign between $J^{\text{NN}}_{\text{Nd–Nd}}$ and $J^{\text{2nd}}_{\text{Nd–Nd}}$ indicates that the RKKY interaction plays the main role in the Nd-Nd interaction. In the calculations, the strong Ising-like anisotropy of the Nd-moments is essential to reproduce the observed data, where $S^{\text{Nd}}$ is found to be smaller than the angular momentum value $J$ (=9/2) of the free Nd$^{3+}$ ions. To understand this results, we consider a doublet with the expectation values of the angular momentum component along the principal axis, $\pm | S^{\text{Nd}} |$ ( $S^{\text{Nd}}$ <$J$), where the strong anisotropy can be expected if the energies of the lowest excited states are much larger than $g_{\text{Nd}}\mu_B \boldsymbol{S}^{\text{Nd}} \cdot \boldsymbol{H}$ and $k_BT$.

Once these values are found, we can determine the behavior of $M_{\text{Nd}}$, $C_{\text{mag}}/T$ and $I_{\text{mag}}$ as well as the magnetic structure of the present system for arbitrary $\boldsymbol{H}$ and $T$. The $H$-dependences of $I_{\text{mag}}$ obtained by fitting the calculation to the observed data (solid circles taken with decreasing $H$) are shown at 1.6 K for $\boldsymbol{H}$//[001] in Figs. 4(a)-4(d), by the solid lines. In the calculation, the intensity average has been taken over the crystallographically equivalent Bragg points in the plane perpendicular to $\boldsymbol{H}$ to take account of the domain structure. We have found that the $H$-dependence of the calculated $I_{\text{mag}}$ can explain the experimental data. The average internal magnetic field at the Nd sites from the Mo-moments $\langle \boldsymbol{H}^{\text{Nd}\leftarrow\text{Mo}}_{\text{int}} \rangle$ is 3.0 T, at which the observed values of $I_{\text{mag}}$ becomes essentially zero at several Bragg points, indicating that the net magnetization of Nd moments $M_{\text{Nd}}$ disappear. Arrows have been also obtained by fitting to the $H$=0 data (shown by the open circles), where the distribution probabilities of the ferromagnetic domains among the three equivalent magnetization directions, [001], [010] and [100] are optimized.

In Figs. 4(e)-4(g), the magnetic correlation patterns whose probabilities are dominant, are shown for the Mo$_4$- and Nd$_4$-tetrahedra at 1.6K under the magnetic field along $\boldsymbol{H}$//[001] at $H$~0, $H$=3 T and $H$=6 T, respectively. The solid arrows indicate the directions of the magnetic moments and the open arrows show the net moment direction of the Nd$^{3+}$ tetrahedra, which has the "two-in two-out" structure of the Nd-moments. The magnetic configuration with the largest probability (and the lowest energy $\varepsilon_i$) at $H$~0 is described as described below: The Mo-moments align along the direction nearly parallel to the [001] direction but have the slight tilting by the angle ~1.0° from the [001] direction towards the direction opposite to the corresponding local principal axes. The Nd moments have the "two-in two-out" configuration with their net magnetization being antiparallel to that of the Mo-moments. This magnetic configuration is reasonable for the values of $J_{\text{Mo–Mo}}$, $J_{\text{Mo–Nd}}$ and $J^{\text{NN}}_{\text{Nd–Nd}}$. The probability of the other magnetic configurations increases with increasing $H$. In the applied magnetic field $H$=3 T, all possible correlation patterns under the restriction of the "two-in two-out" condition are almost degenerated, which is similar to the ground state of spin ice systems,[27,28)] *e.g.* Dy$_2$Ti$_2$O$_7$ and Ho$_2$Ti$_2$O$_7$. In R$_2$T$_2$O$_7$ (T=transition metal elements), the mechanism of the interactions among the rare earth (R) moments seems to depend on the species of R and T atoms: In Nd$_2$Mo$_2$O$_7$, the interaction is by the RKKY mechanism, whereas in Ho$_2$Ti$_2$O$_7$, it is the dipole interaction. Anyhow, there are many correlation patterns, which coexist at $H$~3 T even under the condition of the "two-in two-out" structure. At the applied field of $H$=6 T, the magnetic configuration with the largest probability is described as follows. The Mo-moments align along the direction nearly parallel to $\boldsymbol{H}$//[001], but have the slight tilting by the angle 1.1° from $H$//[001] direction towards the corresponding local principal axes. The Nd moments have the "two-in two-out" configuration with their net magnetization being parallel to that of the Mo-moments.

The results of the $I_{mag}$ calculated for $H$//[0$\bar{1}$1] are shown at 1.6 K against $H$ in Figs. 5(a)-5(e). The magnetic configuration with the largest probability at $H \to 0$ indicates that the Mo-moments align along the direction nearly parallel to $H$//[0$\bar{1}$1], but they are tilted from $H$//[0$\bar{1}$1] and the Nd moments have the "two-in two-out" configuration shown in Fig. 5(f). The magnetic configuration with the largest probability at $H$=6 T indicates that the Nd moments have the "two-in two-out" configuration shown in Fig. 5(h) and the Mo-moments align along the direction nearly parallel to $H$//[0$\bar{1}$1], but they are tilted from $H$//[0$\bar{1}$1] by the Mo-Nd interaction. In the applied field of 3 T, the correlation pattern of the ground state has the two-fold degeneracy, as shown in Fig. 5(g). The $H$-dependence of the magnetic structure is basically same as the one reported previously by authors' group.[6] (In ref. 6, the interaction among the Nd-moments is neglected. However, the ferromagnetic Nd-Nd interaction is effectively considered by adopting the restriction of the "two-in two-out" condition in the calculation.) In the present paper, we have considered all local magnetic structures of 16 Mo- and 16 Nd-moments within the f.c.c. unit cell without the restriction of the "two-in two-out" condition, where the probabilities of these configurations are considered by the canonical distribution. (As stated above, we used the periodic boundary condition.) It is stressed that we have not used the averaged magnetic structure, which may not give the correct results of the behavior of the chirality $\chi$ of the magnetic moments, because $\chi$ is defined locally and should be calculated for the local spin configurations. In this sense, the results we present below present the improved version of those reported in ref. 6.

In Fig. 6, the $T$-dependence of $I_{mag}$ calculated in zero magnetic field for 200 reflection is shown by the solid line together with the observed data. The calculated values of $M_{Nd}$ are shown at various temperatures against $H$ along [001], [110] and [111] in Figs. 7(a)-7(c), respectively. We have found that the calculated data of $I_{mag}$ and $M_{Nd}$ agree quantitatively with the experimentally observed data. In Figs. 8(a) and 8(b), the calculated values of $C_{mag}/T$ are shown as a function of $T$ under the various magnetic field along $H$//[001] and $H$//[111], respectively. The insets of Figs. 8(a) and 8(b) show the observed $C_{mag}/T$–$T$ data in $H$//[001] and $H$//[111], respectively. The characteristics of the calculated $C_{mag}/T$–$T$ curves in the applied field along $H$//[001] and $H$//[111] are also consistent with the experimentally observed data, that is, the peak position, peak width and the $H$- and $T$-dependences are reproduced rather well. The calculated $C_{mag}/T$–$T$ data in $H$=0 can reproduce the experimentally observed data indicating that $\omega_Q$ ($=2\mu_{Nd}\cdot\langle H_{int}\rangle$) and $\Gamma(Q)$ obtained by the inelastic neutron studies can be explained by the calculation. To reproduce the broadened peak width of the $C_{mag}/T$–$T$ curves, it is important to consider the second nearest neighbor Nd-Nd interaction as well as the one between the neighboring Nd-Nd moments. These results indicate that not only the $H$-dependences but also the $T$-dependences of the experimentally observed data can be explained by the present calculation using the material parameters determined in the present work.

The chirality mechanism has been proposed that the ordering of the spin chirality $\chi$ ($\equiv S_1 \cdot S_2 \times S_3$) contributes to the Hall resistivity $\rho_H$.[18,19] According to this mechanism for strongly correlated electrons, the fictitious magnetic flux ($\Phi$) induced by the spin chirality $\chi$ acts on conducting electrons in the same way as the real magnetic field, where $\rho_H$ cannot be treated by the perturbation of the spin-orbit interaction. The fictitious magnetic flux, which an electron feels when it moves the loop drawn by the lines connecting three moments $S_1$, $S_2$ and $S_3$, is proportional to $\chi \equiv S_1 \cdot S_2 \times S_3$ and the fictitious magnetic flux $\phi$ due to the four moments located at the corner of a tetrahedron, is described the expression:

$$\phi = \sum_{\langle l,m,n \rangle} (S_l \cdot S_m \times S_n) \hat{n}_{l,m,n} \quad , \quad (8)$$

where $\hat{n}_{l,m,n}$ is a unit vector perpendicular to a triangle formed by a spin site $l$, $m$ and $n$. The sum is taken over the sets of different numbers of $l$, $m$ and $n$ chosen in a cyclic way from 1-4. The total fictitious magnetic flux $\Phi$ of the system is the vector sum of $\phi$ of all tetrahedra. According to this mechanism, the anomalous Hall conductivity $\sigma_H (= -\rho_H/\rho^2)$ is proportional to the parallel component of the fictitious magnetic flux to the external magnetic field, $\Phi^{//}$. For Nd$_2$Mo$_2$O$_7$, the Hall conductivity $\sigma_H$ is nearly proportional to the Hall resistivity $\rho_H$, because the electrical resistivity $\rho$ below ~60 K is almost $T$- and $H$-independent in the $H$-region of $H \leq 7$ T.[11,29] Then, $\rho_H$ of Nd$_2$Mo$_2$O$_7$ is expected to be proportional to the component parallel to $H$ contributed from the Mo-moments ($\Phi_{Mo}^{//}$). We have calculated $\Phi_{Mo}^{//}$ of Nd$_2$Mo$_2$O$_7$ as a function of $H$ and $T$ by using the obtained material parameters derived above, where $\Phi_{Mo}^{//}$ is estimated from the averaged fictitious flux generated by the local chirality, $\chi=\langle S_1 \cdot S_2 \times S_3 \rangle$, not by the one described by $\chi=\langle S_1 \rangle \cdot \langle S_2 \rangle \times \langle S_3 \rangle$. Specifically, the values of $\Phi_{Mo}^{//}(H,T)$ are calculated by using expression as follows:

$$\Phi_{Mo}^{//}(H,T) = \frac{\sum_i \sum_{\langle l,m,n \rangle} \left[ (S_{l,i}^{Mo} \cdot S_{m,i}^{Mo} \times S_{n,i}^{Mo}) \hat{n}_{l,m,n} \cdot \hat{H} \right] \exp(-\varepsilon_i/(k_B T))}{\sum_i \exp(-\varepsilon_i/(k_B T))},$$
(9)

where $\varepsilon_i$ and $S_{l,i}^{Mo}$ are the energy and the spins at the $l$–th ($l$=1-16) Mo site, respectively, for the $i$-th magnetic configuration of the Nd-moments. The sum is taken over all sets of different numbers of $l$, $m$ and $n$ chosen from 1-16 under the condition that these three sites form a single triangle surface of a Mo$_4$-tetrahedron. The probability of the $i$-th magnetic configuration can be calculated as a function of $\varepsilon_i$ by using the canonical distribution.

In Figs. 9(a)-9(c), the calculated fictitious flux components parallel to $H$ contributed from the Mo-moments ($\Phi_{Mo}^{//}$) are shown at 5 K against $H$ along [001], [110] and [111], respectively, together with the observed $\rho_H$ data, where the calculated values are scaled to coincide with the observed ones. According to the chirality mechanism, the $H$-dependences of $\Phi_{Mo}^{//}$ should agree with those of the observed $\rho_H$ data. However, we cannot find the agreement in the figures. We note here that the ordinary Hall resistivity is not significantly large and its consideration does not change the arguments. For $H$//[001], for example, the observed $\rho_H$ decreases monotonically with increasing $H$, while the calculated $\Phi_{Mo}^{//}$ has a minimum at around $H$= 3.0 T, where the exchange field at Nd sites is canceled by the applied field. This characteristic behavior of the calculated curve can roughly be understood as follows. Because $\Phi_{Mo}^{//}$ of the Mo-moments induced by the interaction with the Nd-moments, which form, for example, the "two-in two-out" structure with the net moment along the $x$- or $y$-axis, is very much smaller than that for the case of the net moment along the $z$-axes, the $\Phi_{Mo}^{//}$-$H$ curve has the minimum at $H$=3.0 T. (Consider that at $H$=3 T, the probability of the "two-in two-out" structure with the net moment along the $x$- or $y$-axis becomes maximum, as shown in Fig. 4(f). )

For $H$//[111], the observed $\rho_H$ decreases rather rapidly with increasing $H$, while the calculated $\Phi_{Mo}^{//}$ does not exhibit such the rapid decrease. Moreover, the $\rho_H$-$H$ curve is concave, while the $\Phi_{Mo}^{//}$-$H$ curve exhibits the convex behavior. The behavior of the $\Phi_{Mo}^{//}$-$H$ curve is naturally understood by considering the $H$-dependence of the magnetic structures with relatively large probabilities: They are shown at 5 K under the magnetic field along [111] for $H\to 0$, $H\sim 3$ T, and $H$=6 T in Figs. 10(a)-10(c), respectively, where the thick arrows indicate the directions of the magnetic moments at the corresponding sites and the open arrows show the net moment direction of the $Nd^{3+}$ tetrahedra, which has the "two-in two-out" structure of the Nd-moments. In the figure, the signs of $\Phi_{Mo}^{//}$ induced by the magnetic interaction of Mo moments with the Nd moments in the lower two rows, are indicated for various Nd ordering patterns. At $H\to 0$, the probability is relatively large for the "two-in two-out" and "one-in three-out" structures as shown in Fig. 10(a). With increasing $H$, the probability of the "one-in three-out" structure decreases and that of the "two-in two-out" structure becomes dominant as shown in Fig. 10(b). With further increasing $H$, the probability of the "three-in one-out" structure shown in Fig. 10(c) increases. The drastic change of the probability of the magnetic configuration originates from the competition among the Mo-Nd and Nd-Nd interactions and Zeeman energy. As shown in the figure, the sign of $\Phi_{Mo}^{//}$ is positive, if the Nd-moments have the "two-in two-out" structure, while the sign of $\Phi_{Mo}^{//}$ is negative if the Nd-moments have the "one-in three-out" and "three-in one-out" structures. Then, because the latter two structures are slightly mixed at $H\to 0$ and $H$=6 T, the calculated $\Phi_{Mo}^{//}$-$H$ curve exhibits a slight decrease as $H$ deviates from $\sim$3 T, which can explain the convex behavior of the calculated line. It is not experimentally observed.

In Figs. 11(a) and 11(b), the calculated $\Phi_{Mo}^{//}$ at 1.0 K is plotted against the magnetic fields $H$ applied along [001] and [111], respectively, together with the $\rho_H$ data[11] taken at various temperatures with the magnetic field up to 15 T. The curves of $\Phi_{Mo}^{//}$ do not agree with experimentally observed $\rho_H$. Taguchi et al. argued that the sign change of $\rho_H$ with increasing $H$ along [111] was an evidence that the chirality mechanism is actually working in the present system.[20] However, the $H$-dependence of $\Phi_{Mo}^{//}$ is quite different from that of the observed $\rho_H$-$H$ curve.

Next, the $T$-dependence of $\Phi_{Mo}^{//}$ is discussed. We calculate the fictitious flux by assuming that even at $H$=0, $\Phi_{Mo}$ are parallel to [001], that is, the ferromagnetic domains are aligned. In Fig. 12, $\Phi_{Mo}^{//}$ calculated for $H$=0 is shown against $T$. The $T$-dependence of the observed $\rho_H$ taken under the relatively weak magnetic field $H$=0.5 T along $H$//[001] is also plotted in Fig. 12. Because at $H$=0.5 T, the magnetic domains are aligned but the magnetic structure is not significantly changed from that at $H$=0, the observed $\rho_H$ can be compared with the calculated results at $H$=0. We find, however, that the $T$-dependence of $\Phi_{Mo}^{//}$ is quite different from that of the observed Hall resistivity as shown in Fig. 12. In the above calculations of the fictitious magnetic flux, we have estimated from the averaged flux defined by the local arrangement of the moments not by the averaged structure of the ordered moments. The finite $\Phi_{Mo}^{//}$ are found in the $T$-region $T$>30 K where the ordering of the Nd-moments practically disappear. However, the $T$- and $H$-dependences of the fictitious magnetic flux deduced from the obtained material parameters do not agree with those of the observed Hall resistivity $\rho_H$ of $Nd_2Mo_2O_7$, indicating that the unusual behavior of $\rho_H$ is not explained only by the chiral order mechanism.

The calculated results of the Hall resistivity due to the chirality mechanism are reported in Ref. 25, where the value of $J_{Mo-Nd}$ =-2.7 K was adopted to reproduce the $T$-dependence of the observed $\rho_H$ of $Nd_2Mo_2O_7$. (They estimated the value of $J_{Mo-Nd}$ =-2.7 K from $T^*$ ~50 K.) The value of $J_{Mo-Nd}$ =-2.7 K is much larger than the value obtained by the presented study ($J_{Mo-Nd}$ =-0.24 K). It was reported that the tilting angle from the [001] direction of the Mo-moments is estimated to be ~4° at 2 K in $H$=0 in order to explain the absolute value of $\rho_H$ by the chirality mechanism.[19] The tilting angle ~4 ° which is larger than the obtained value by the presented study (=1.0 °), indicates that they overestimated the Mo-Nd interaction.

To clarify the relationship between the chirality mechanism and the unusual behavior of the Hall resistivity, we have analyzed the magnetic structure in detail and based on the knowledge, we have reached the conclusion that the $T$- and $H$-dependences of the fictitious magnetic flux deduced from the obtained material parameters cannot explain the observed behavior of the Hall resistivity of $Nd_2Mo_2O_7$ not only quantitatively but also qualitatively.

The behavior of $\rho_H$ for $Nd_2Mo_2O_7$ is described by the phenomenological equation $\rho_H$=$R_0 H$+$4\pi R_s M_{Mo}$+$4\pi R_s$'$M_{Nd}$, where two components of anomalous Hall resistivity are necessary. However, the anomalous Hall resistivity component of the Nd-moments cannot be explained by any conventional mechanism as well as the chirality one. We have found that the RKKY interaction plays the important role in the Nd-Nd interaction. It is naturally considered that the spin of the conduction electrons of Mo atoms polarized by the interaction with the Nd-moments contribute to the anomalous Hall resistivity. Possible specific explanation of the behavior $\rho_H$ for $Nd_2Mo_2O_7$ by the mechanism still remains a future issue.

In summary, we have shown the neutron scattering data taken for a single crystal of $Nd_2Mo_2O_7$ in the applied magnetic field $H$//[0$\bar{1}$1] and $H$//[001] up to 5.7 T as well as inelastic neutron scattering data taken by using cold neutrons. We have determined the material parameters such as the single ion anisotropies of the Mo- and Nd-moments and exchange coupling constants between the Mo-Mo, Mo-Nd and Nd-Nd magnetic moments to reproduce the neutron data, magnetization curve and the behavior of the magnetic specific heat simultaneously. In the analyses, we have to consider not only the nearest neighbor Nd-Nd interaction but also the second nearest neighbor one, indicating that the RKKY interaction has a certain role. From the obtained material parameters, the spin chirality and the fictitious magnetic flux $\Phi_{Mo}$ have been calculated as a function of $H$ and $T$ by using the chirality mechanism, where $\Phi_{Mo}$ is estimated from the averaged fictitious flux generated by the local chirality, $\chi=<S_1 \cdot S_2 \times S_3>$, not by the one described as $\chi=<S_1> \cdot <S_2> \times <S_3>$. Comparison of the Hall resistivity $\rho_H$ deduced from $\Phi_{Mo}$ with the experimentally observed data, indicates that the chiral order mechanism is not, at least, playing a main role in determining the observed behavior of $\rho_H$ in the present system.

**Acknowledgments**

The authors thank Professor Hirashima of Nagoya University for discussion. We also thank the Neutron Science Laboratory of the Institute for Solid State Physics (NSL-ISSP) for the use of the instrument within the national user's program. The work is supported by Grants-in-Aid for Scientific Research from the Japan Society for the Promotion of Science (JSPJ) and by Grants-in-Aid on priority areas from the Ministry of Education, Culture, Sports, Science and Technology.

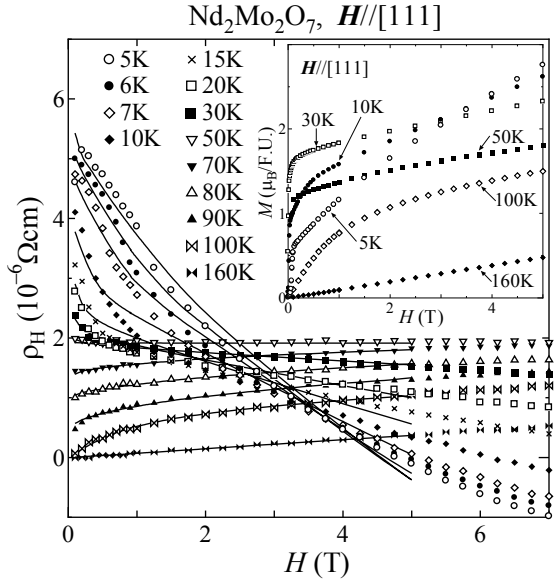

Fig. 1. Hall resistivity $\rho_H$ of $Nd_2Mo_2O_7$ is shown at various temperatures against the magnetic field $H$ applied along [111]. Solid lines show the results of the fittings. Details of the fittings are described in the text. Inset shows the magnetization curves of $Nd_2Mo_2O_7$ measured under the applied magnetic field along the [111] direction. The Hall resistivity and magnetization data were reported previously by the authors' group.[7,8]

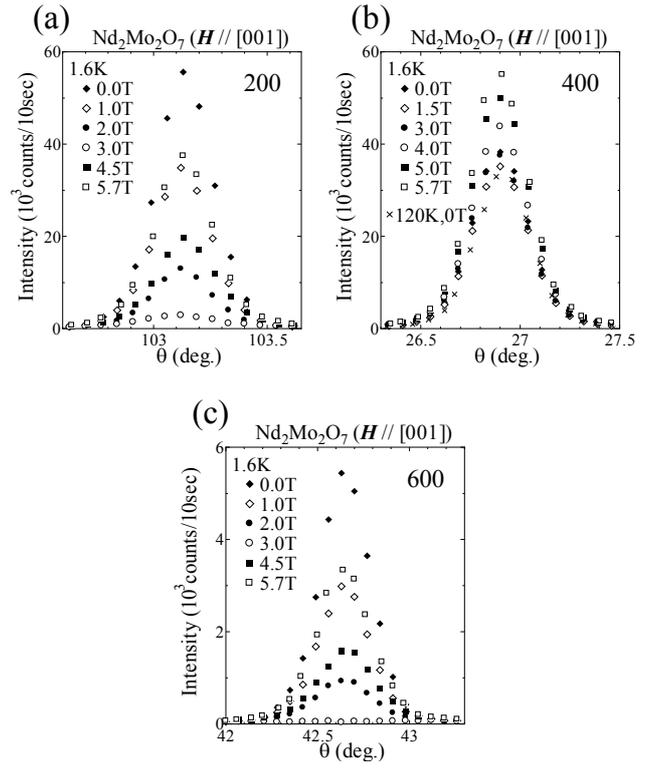

Fig. 3. Profiles of the ω–scans (sample-angle scans) for 200, 400 and 600 reflections taken under the magnetic filed $H$ ($H$ //[001]), are shown in (a)-(c), respectively. The values of $H$ and $T$ are shown in the figures.

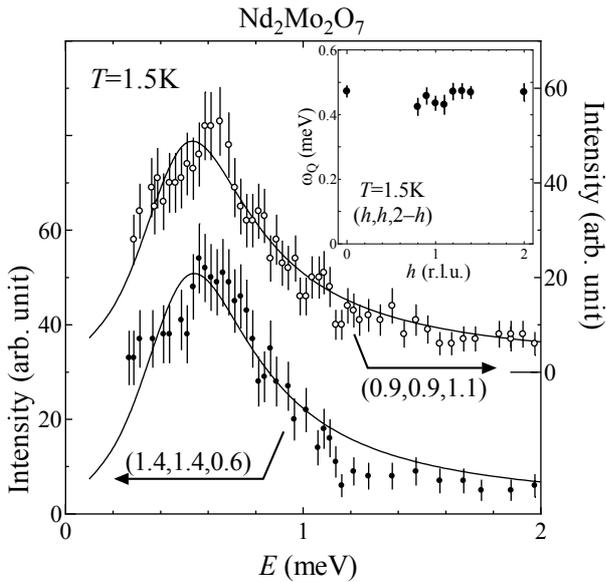

Fig. 2. Energy-scan profiles taken at 1.5 K at $Q$=(0.9,0.9,1.1) and (1.4,1.4,0.6) with the resolution of $\Delta E$=0.052 meV. The solid lines show the results of the fittings. Inset shows the dispersion curve along ($h, h, 2-h$) in the reciprocal space.

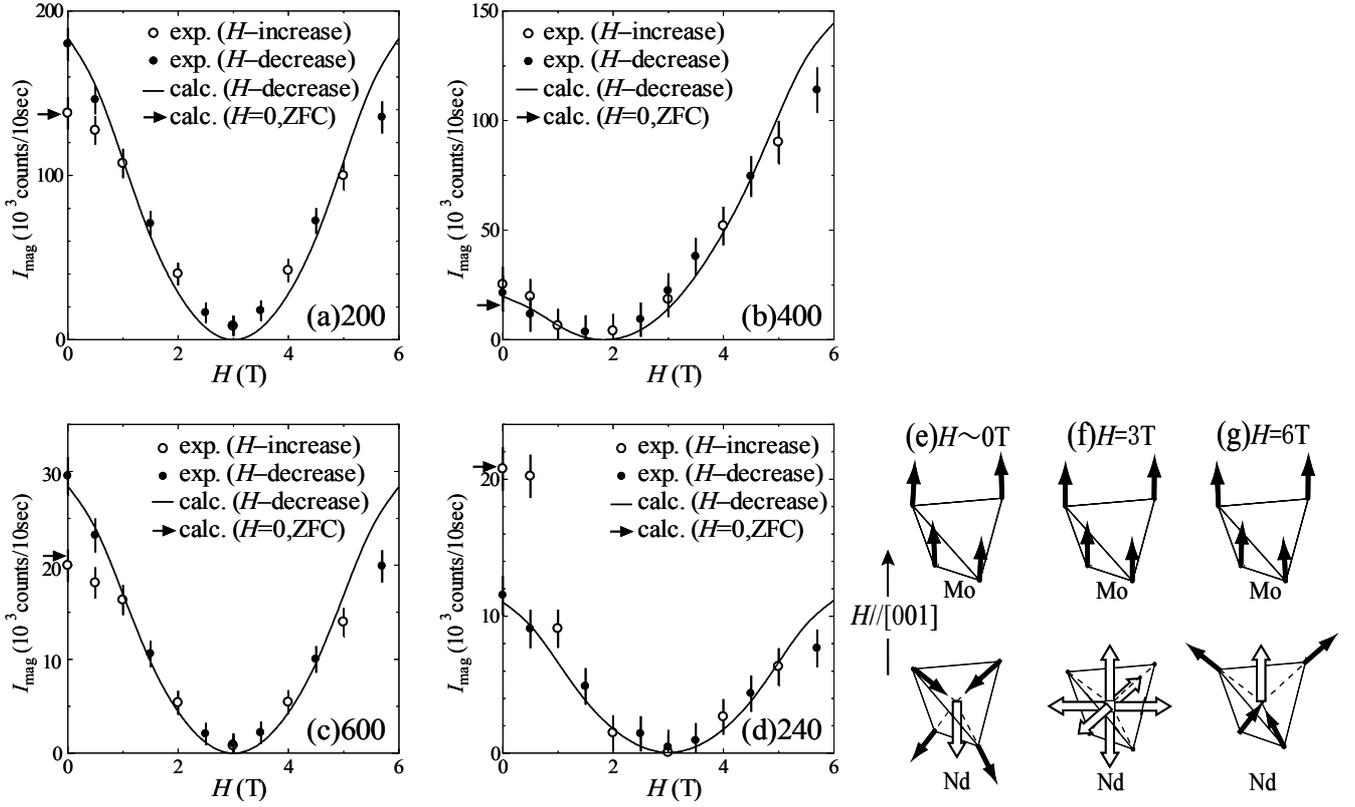

Fig. 4. The integrated intensities of the magnetic scattering Nd$_2$Mo$_2$O$_7$ at 1.6 K are shown against the magnetic field $H$ (//[001]) for 200, 400, 600 and 240 reflections in (a)-(d), respectively. These data were taken after the zero field cooling (ZFC) by scanning $H$ stepwise up to 5.7 T ($H$-increase) and then down to zero ($H$-decrease). Solid lines and arrows in (a)-(d) show the results of the calculation. Details of analysis are described in the text. The magnetic correlation patterns whose probabilities are dominant, are shown for the Mo$_4$- and Nd$_4$- tetrahedra at 1.6K under the magnetic field along $H$//[001] at (e)$H$~0, (f)$H$=3 T and (g)$H$=6 T, respectively. The solid arrows indicate the directions of the magnetic moments. The open arrows show the net-moment directions of the Nd$^{3+}$ tetrahedra with the "two-in two-out" structure. In the applied field of 3T, the ground state of the correlation pattern has the six-fold degeneracy.

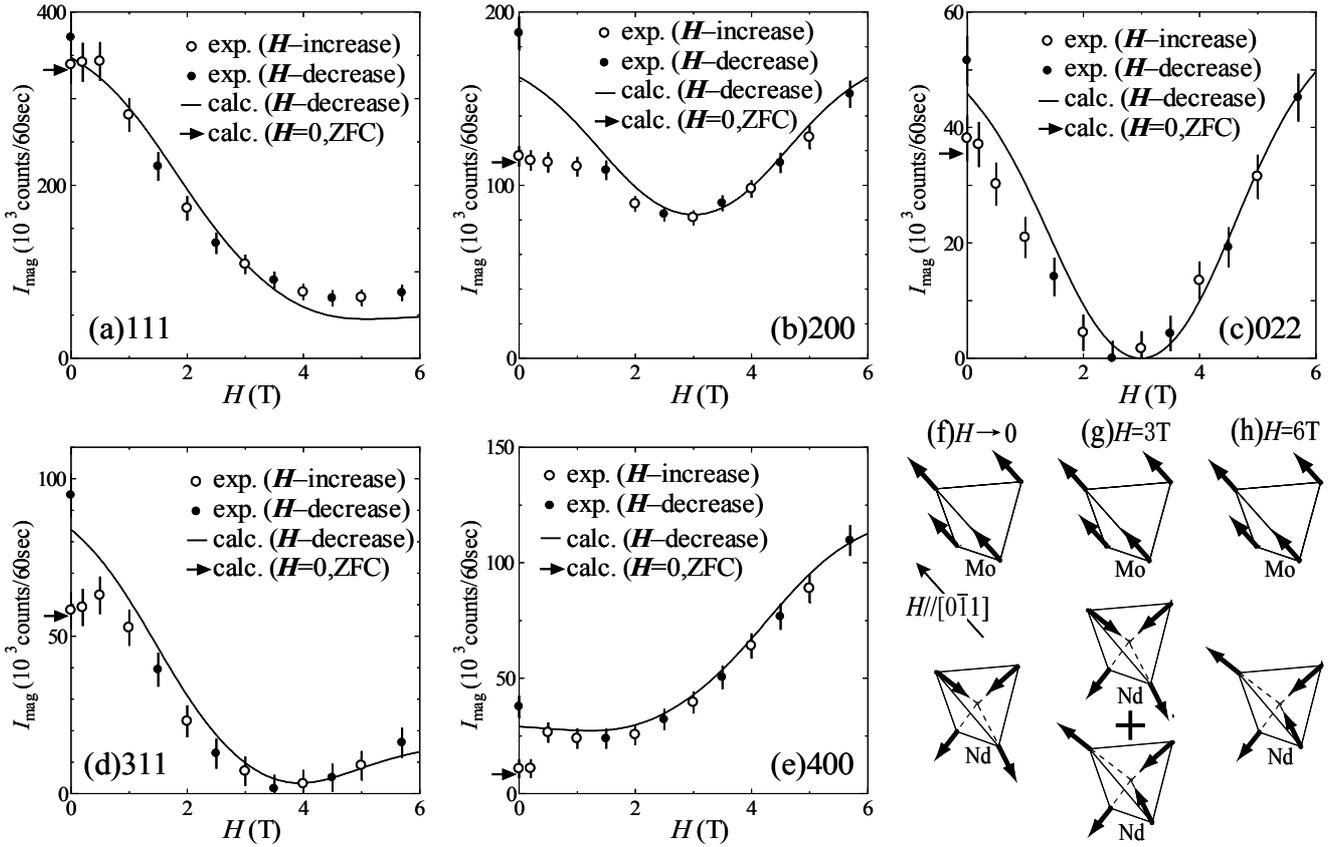

Fig. 5. The integrated intensities of the magnetic scattering of $Nd_2Mo_2O_7$ at 1.6 K is shown against the magnetic field $H$ ($//[0\bar{1}1]$) for 111, 200, 022, 311 and 400 reflections in (a)-(e), respectively. These data were taken after the zero field cooling (ZFC) by scanning $H$ stepwise up to 5.7 T ($H$-increase) and then down to zero ($H$-decrease). The data were reported previously by the authors' group.[6] The solid lines and arrows in (a)-(e), show the results of the calculation. Details of the analysis are described in the text. The magnetic correlation patterns whose probabilities are dominant, are shown for the $Mo_4$- and $Nd_4$- tetrahedra at 1.6K under the magnetic fields along $H//[0\bar{1}1]$ at (f)$H\to 0$, (g)$H$=3 T and (h)$H$=6 T, respectively. The thick arrows indicate the directions of the magnetic moments. In the applied field of 3 T, the ground state of the correlation pattern has the two-fold degeneracy.

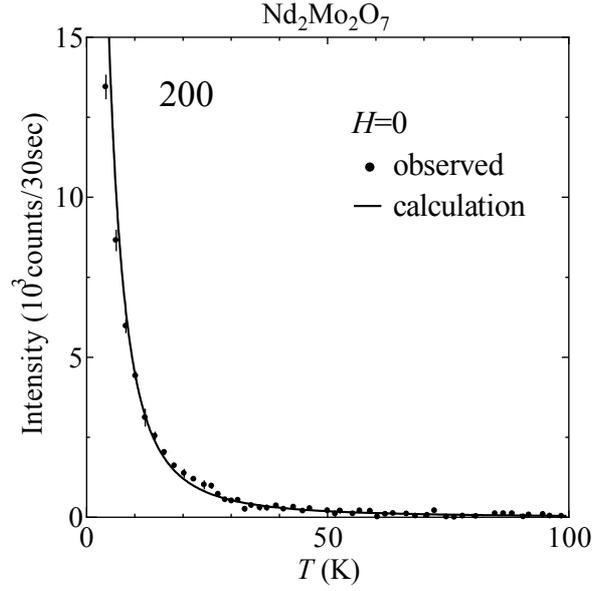

Fig. 6. Temperature dependence of the magnetic scattering intensities of the 200 reflection of $Nd_2Mo_2O_7$ is shown at $H$=0. The solid line shows the calculated intensity.

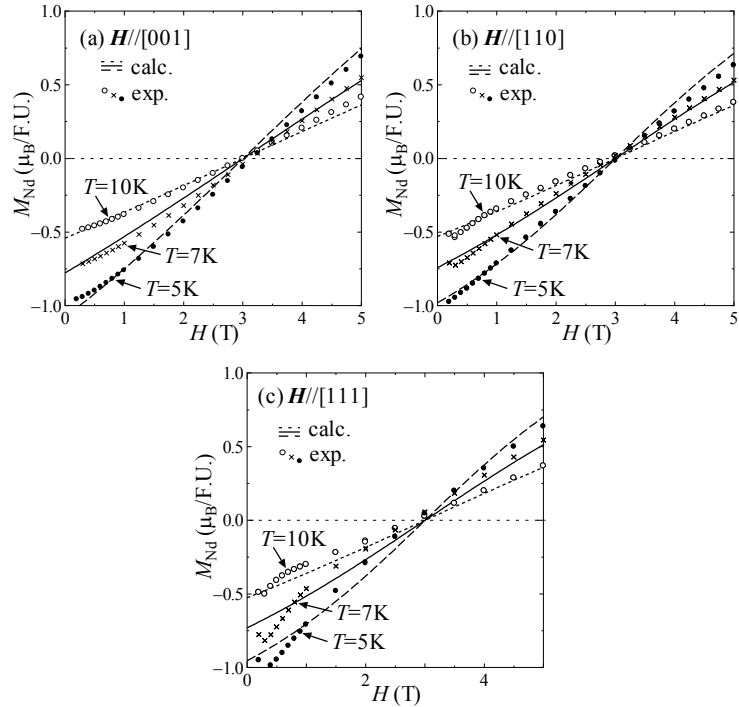

Fig. 7. The Nd-components of the magnetizations ($M_{Nd}$) are shown at various temperatures against the magnetic fields $H$ applied along (a)[001], (b)[110] and (c)[111], respectively, where the $M_{Nd}$ is estimated by subtracting the magnetization at 40 K from those at various temperatures. Parts of the raw data of the magnetization were reported by the authors' group in refs. 6-8. Dashed, solid and dotted lines show the calculated $M_{Nd}$ at 5 K, 7 K and 10 K, respectively.

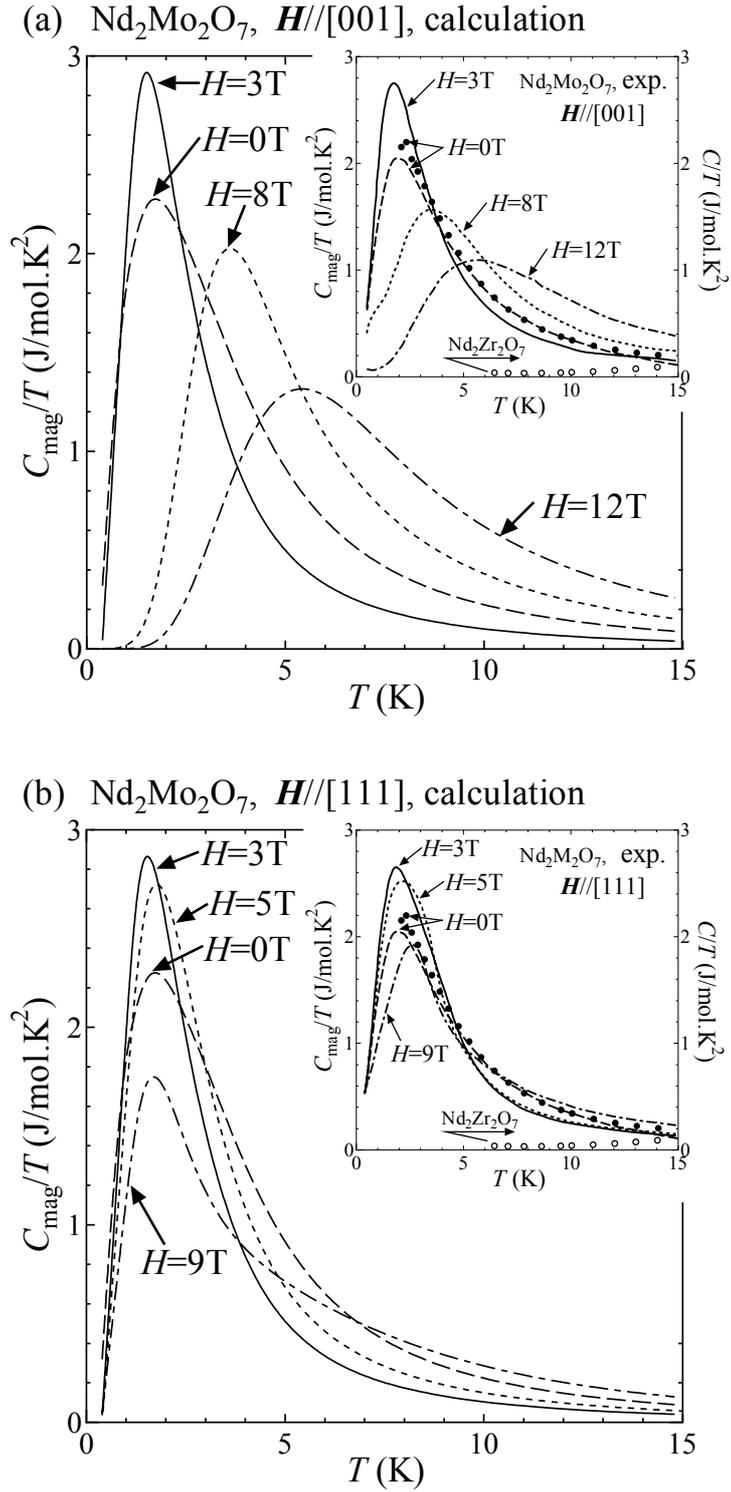

Fig. 8. (a) Calculated specific heat of the magnetic contribution divided by the temperature, $C_{mag}/T$ of $Nd_2Mo_2O_7$ is shown as a function of $T$ under the various magnetic field along $H//[001]$. Inset shows the observed $C_{mag}/T$ data of $Nd_2Mo_2O_7$ (lines) and the $C/T$ data of $Nd_2Zr_2O_7$ (open circles), respectively. The raw data of $C/T$ shown by the solid circles were reported previously by authors' group in refs. 23 and data shown by the lines were reported in ref. 24. (b) Calculated specific heat of the magnetic contribution divided by the temperature, $C_{mag}/T$ of $Nd_2Mo_2O_7$ is shown as a function of $T$ under the various magnetic field along $H//[111]$. Inset shows the observed $C_{mag}/T$ data of $Nd_2Mo_2O_7$ (lines) and the $C/T$ data of $Nd_2Zr_2O_7$ (open circles), respectively. The raw data of $C/T$ shown by the solid circles were reported previously by authors' group in refs. 23 and data shown by the lines were reported in ref. 24.

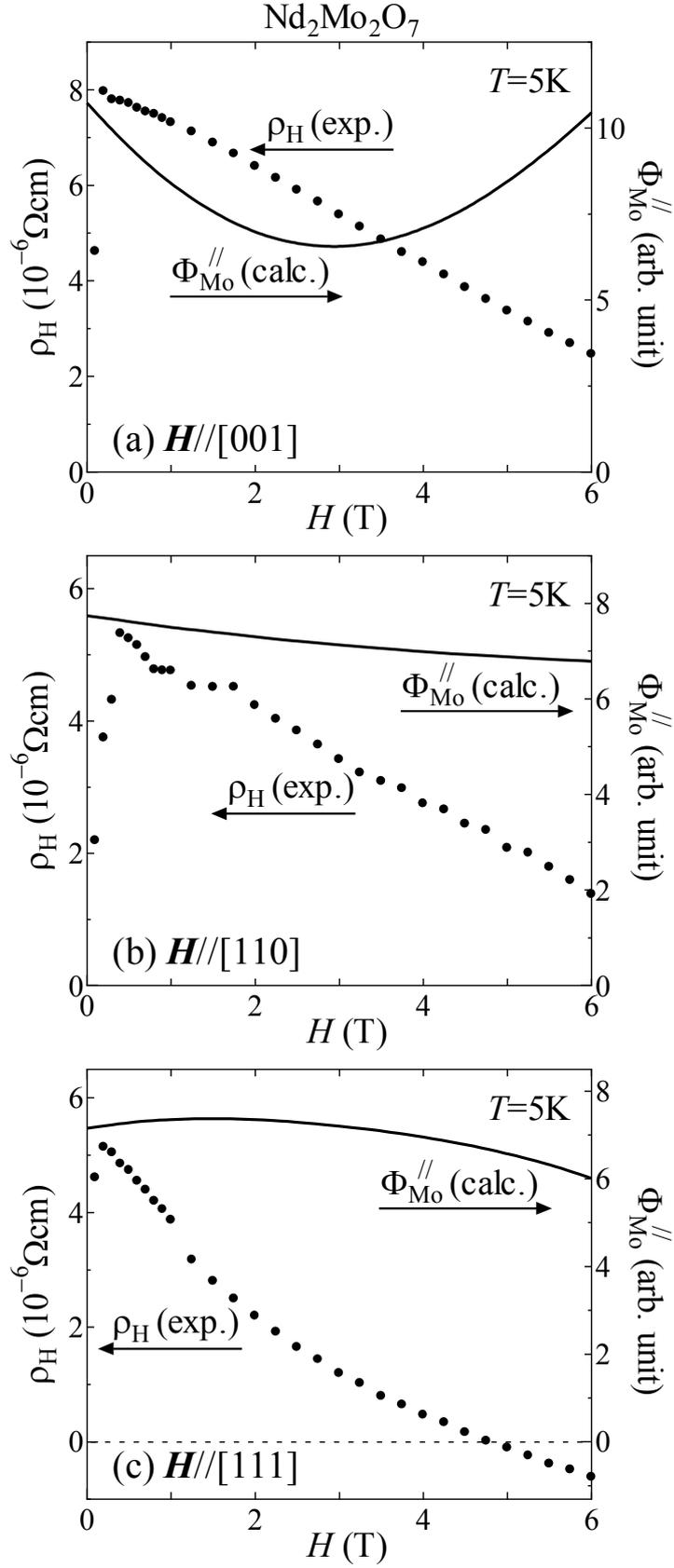

Fig. 9. Calculated fictitious magnetic flux components parallel to $H$ contributed from the Mo-moments, $\Phi_{Mo}^{//}$, are shown at 5 K against the magnetic fields $H$ applied along (a) [001], (b) [110] and (c) [111], respectively. The observed data of $\rho_H$ are shown by the solid circles.

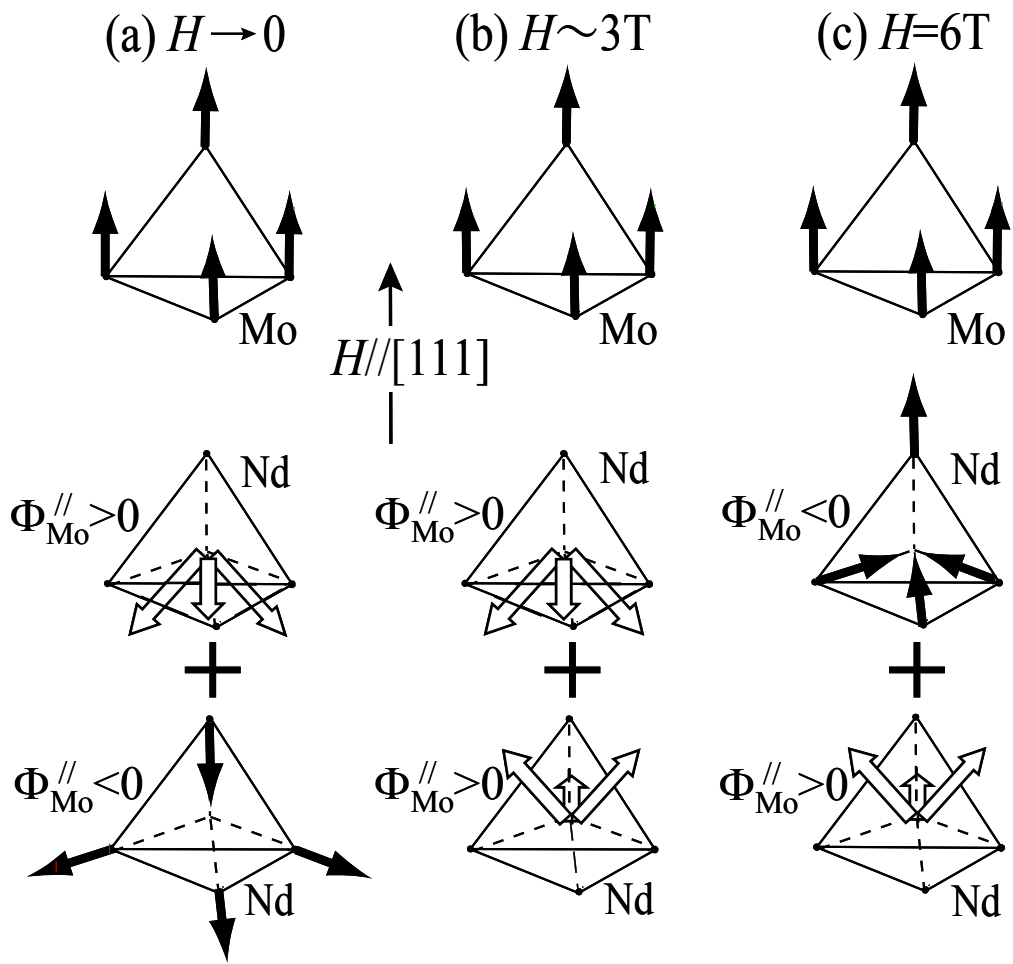

Fig. 10. The magnetic correlation patterns whose probabilities are dominant, are shown for the $Mo_4$- and $Nd_4$- tetrahedra at 5 K under the magnetic fields along ***H***//[111] for (a)$H\to 0$, (b)$H\sim 3$ T and (c)$H=6$ T, respectively. The solid arrows indicate the directions of the magnetic moments. The open arrows show the net moment direction of the $Nd^{3+}$ tetrahedra, which has the "two-in two-out" structure of the Nd-moments. The several different correlation patterns shown in the figures have the almost degenerate energies.

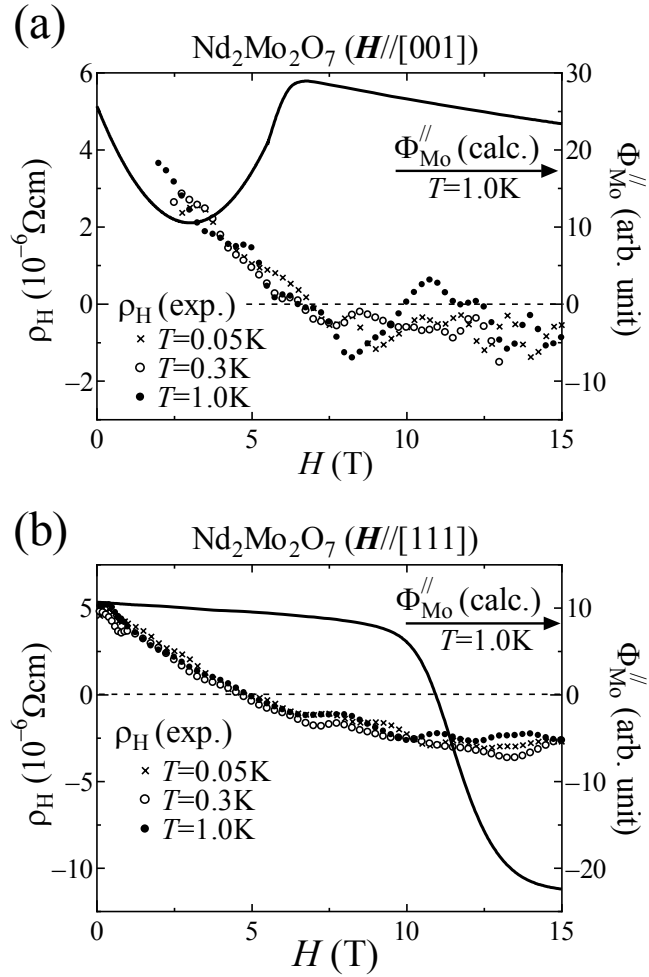

Fig. 11. Calculated $\Phi_{Mo}^{//}$ at 1.0 K is plotted against the magnetic fields $H$ applied along (a) [001] and (b) [111], respectively, together with the observed $\rho_H$ data[11] taken at various temperatures with the magnetic fields up to 15 T.

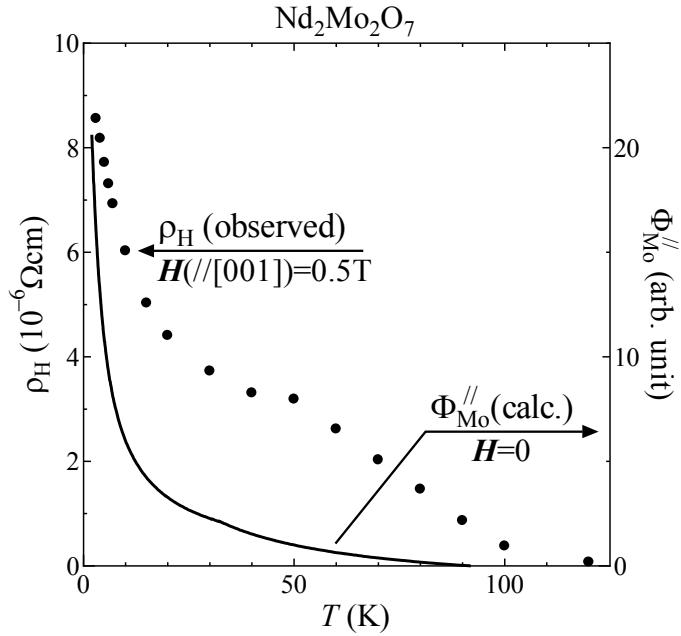

Fig. 12. Calculated fictitious magnetic flux component parallel to the [001] direction contributed from the Mo-moments is shown at $H=0$ as a function $T$. The $\rho_H$ data taken under the magnetic field $H=0.5$ T along the [001] direction are shown by the closed circles.